\documentclass[11pt]{article}
\usepackage{graphicx,psfrag,amsmath,amssymb,amsfonts,bbm,latexsym,color,dcolumn}

\title{\bf Advances in decoherence control}

\author{ {\sc LORENZA VIOLA} \\
{\small \it Los Alamos National Laboratory, Los Alamos, NM 87545, USA,}\\
{\small \it and Department of Physics and Astronomy, 
6127 Wilder Laboratory, } \\
{\small \it Dartmouth College, Hanover, NH 03755, USA }}

\begin{document}

\maketitle

\abstract{ I address the current status of dynamical decoupling 
techniques in terms of required control resources and feasibility.
Based on recent advances in both improving the theoretical design and 
assessing the control performance for specific noise models, I argue 
that significant progress may still be possible on the road of 
implementing decoupling under realistic constraints.   }

\section{Introduction}

Achieving decoherence control has become a critical goal for a variety
of applications in contemporary physics and engineering.  Broadly
speaking, the challenge is to preserve quantum coherence during both
the storage and the manipulation of quantum states in systems which
are unavoidably exposed to the influence of their surrounding
environment -- therefore are subject to irreversible, {\em
open-system} dynamics~\cite{breuer, benatti}.  A number of practical
and conceptual motivations have contributed to make this challenge
increasingly timely and important in recent years.  On one hand,
ensuring a sufficiently high degree of control over both environmental
and operational errors is a chief requirement for physical
realizations of quantum information, shared to a lesser or greater
extent by the large variety of device technologies which are nowadays
being considered~\cite{nc,fort}.  In a broader context, maintaining
coherent quantum behavior in the presence of realistic noise sources
is essential for guaranteeing proper performance by any nanoscale
process or device intended to operate in a quantum-limited
regime~\cite{bm}.  On the other hand, from a more conceptual
standpoint, the development and characterization of control-theoretic
notions and tools for generic open-system evolutions is a fascinating
area {\em per se}, which remains only partially explored to
date~\cite{lv,altafini}.

As a result, the field of decoherence control has rapidly grown
recently, and several quantum stabilization strategies have been
developed, both specifically tailored for quantum information
protection and processing, and beyond.  These include passive
stabilization schemes inspired to quantum reservoir
engineering~\cite{pcz}, decoherence-free-subspace~\cite{dfs} and
noiseless-subsystem coding~\cite{klv} (including topological
approaches~\cite{kita,zl}), as well as active stabilization techniques
based on quantum feedback control~\cite{wise}, quantum
error-correcting codes~\cite{shor,steane}, and dynamical decoupling
methods~\cite{vl,vkl}.  While a discussion of the principles
underlying these different approaches is beyond the present purposes,
I will focus here on critically reconsidering the control resources
involved in the standard dynamical decoupling setting, and argue that
substantial progress might be possible by both further improving the
control design and by better incorporating available knowledge about
the relevant environmental interaction.

\section{Dynamical decoupling framework}

Dynamical decoupling methods for open quantum systems derive their
basic physical intuition from coherent averaging techniques in
high-resolution nuclear magnetic resonance (NMR)
spectroscopy~\cite{hahn, haeberlen}, decoupling being thereby
understood as the selective removal of unwanted contributions to the
nuclear spin Hamiltonian via the application of suitable pulse
sequences.  In the current control-theoretic terminology, decoupling
is usually interpreted in a broader sense than the original one,
see~\cite{vl,vkl,vlk,z,vt,v,lw,bl,vk} for recent representative
literature.  A definition which I will adopt within the present
discussion is to generally understand a decoupling scheme as an {\em
open-loop} dynamical control protocol which relies on the repeated
application of pulsed or switched controls drawn from a {\em finite}
set in order to attain a desired control objective.  In what follows,
I will focus on the prototypical application of decoupling, in which
case the objective is the {\em averaging} of the environmental
couplings responsible for decoherence.  In general, however,
alternative control scenarios may be envisaged, notably the synthesis
of controlled evolutions according to predetermined symmetry
criteria~\cite{vkl,z}, and Hamiltonian quantum simulations~\cite{simu}.

The relevant decoupling setting may be pictorially schematized as 
in figure 1.  The joint evolution of the target system $S$ in interaction
with the environment $E$ is described by a total drift Hamiltonian
of the form
\begin{equation}
H_0=H_S \otimes {\mathbb I}_E + {\mathbb I}_S \otimes H_E + H_{SE}\:,
\hspace{3mm}H_{SE}=\sum_a S_a \otimes E_a \:,
\label{open}
\end{equation}
where $H_S$ and $H_E$ account for the isolated dynamics of the 
system and the environment, respectively, and the interaction term 
$H_{SE}$ is responsible for introducing unwanted decoherence and 
dissipation effects in the reduced dynamics of $S$ alone.  The 
basic idea is to adjoin a judiciously designed controller, described 
by a classical time-dependent field $H_c(t)$ acting {\em only} on the 
system, in such a way that the resulting controlled dynamics 
is described by an effective Hamiltonian $H_{eff}$ which 
{\em no} longer contains mixing terms between $S$ and $E$ that is, 
\begin{equation}
H_0 \mapsto H_0 + H_c(t) \; \text{\ such that \ } 
H_{eff} = \tilde{H}_S \otimes {\mathbb I}_E + {\mathbb I}_S 
\otimes H_E \:,
\end{equation}
for an appropriate, possibly modified, system Hamiltonian 
$\tilde{H}_S$.

\begin{figure}[t]
\begin{center}
\includegraphics[width=4.5in]{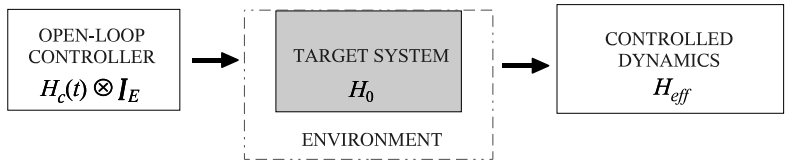}
\end{center}
\caption{ {\small Block-diagram of an open-loop controlled quantum system,
$S$, in interaction with an environment, $E$.  {\em No} direct control 
action on the latter is assumed. } }
\end{figure}

Decoupling protocols are most conveniently constructed by directly
looking at the control propagator associated to $H_c(t)$ as the basic
object for control design, 
\begin{equation}
U_c(t) =\text{T}\exp\left\{ -i \int_0^t  dx \,H_c(x) \right\} \:,
\end{equation}
where T denotes time ordering and units such that $\hbar=1$ have
been chosen.  The analysis and complexity of the control problem are
influenced by various factors.  Beside the available knowledge 
about the target system and the relevant noise generators $S_a$, 
a critical role is played by the assumptions on the accessible
control resources -- in particular, bounded vs unbounded control 
strengths and rates, and/or faulty control operations as opposed 
to perfect ones.

\section{Bang-bang dynamical decoupling}
\subsection{The simplest setting}

One can gain a concrete feeling on how the above strategy works by
revisiting the simplest decoherence control scenario, namely 
a single spin-1/2 system (a qubit) diagonally coupled to an harmonic 
reservoir, 
\begin{equation}
H_0 = \omega_0 \sigma_z \otimes {\mathbb I}_E + {\mathbb I}_S \otimes
\sum_k \omega_k b^\dagger_k b_k + \sigma_z \otimes \sum_k g_k 
(b^\dagger_k + b_k) \:, 
\label{sb}
\end{equation}
where $\omega_0, \omega_k, g_k$ are real parameters,
$\sigma_z=\text{diag}[+1,-1]$ is the $\hat{z}$-Pauli matrix, and
$b^\dagger_k, b_k$ are bosonic creation and annihilation operators.
Since the interaction term $H_{SE}$ has opposite sign depending on
whether the spin is up ($|0\rangle$) or down ($|1\rangle$), one may
intuitively expect that it should be possible to average such
contributions to zero by flipping the qubit {\em faster} than the
typical time it takes to the environment oscillators to appreciably
evolve.

Formalizing such intuition has led to the first example of a {\em
bang-bang} (BB) decoupling protocol~\cite{vl}.  In this case, the
appropriate control action is implemented by a train of resonant,
infinitely short and strong (BB) $\pi$-pulses about the $\hat{x}$ (or
$\hat{y}$) axis, separated by a time interval $\Delta t$.  In terms of
the relevant control propagator $U_c(t)$, notice that this corresponds
to a {\em cyclic} action with a period $T_c =2 \Delta t$, that is,
$U_c(T_c)= (\hat{\pi})^2= {\mathbb I}_S$, with $\hat{\pi}=i\exp(-i \pi
\sigma_x/2)$.  The controlled evolution of the qubit coherence
element, $\rho_{01} (t)$, may be evaluated explicitly in this case. In
the limit of {\em arbitrarily fast control}, this leads to the exact
result
\begin{equation}
\lim_{T_c \rightarrow 0, N \rightarrow \infty }
\rho_{01} (t= N T_c) = \rho_{01} (t=0) \:,
\label{suppr}
\end{equation}
implying, in principle, the possibility to {\em stroboscopically} preserve
quantum coherence indefinitely in time. In practice, the essential physical
requirement making decoherence suppression possible is the ability to access
control timescales $T_c$ over which the system-environment coupling can be 
{\em coherently} manipulated.  Accordingly, if $\tau_c$ denotes the shortest
correlation time associated with the reservoir dynamics, the condition 
$T_c \ll \tau_c$ is always necessary and sufficient in order to approach
the ideal limit of Eq. (\ref{suppr}) where decoherence is {\em completely}
inhibited.  For the linear spin-boson model of Eq. (\ref{sb}), $\tau_c$ is 
essentially determined by the inverse of the highest frequency component
present in the oscillator bath.  

By analogy with the well-know Carr-Purcell (CP) $\pi$-pulse sequence
which is routinely used in NMR to remove unwanted phase
evolutions~\cite{ernst}, I will refer to this simple single-qubit
decoupling protocol as CP-decoupling (see also~\cite{speranta} for an
explicit comparison between standard and time-symmetric CP-schemes in
this model and~\cite{chi} for the extension to nonlinear spin-boson
couplings).  Interestingly, a proof-of-principle demonstration of
CP-decoupling to suppress single-photon decoherence in an unbalanced
Michelson polarization interferometer has been reported
in~\cite{berglund}.  Variants of this basic scheme (including
$2\pi$-pulse protocols that are related to $\sigma_z$-decoupling) have
found application in problems as different as suppression of
collisional decoherence~\cite{search}, inhibition of spontaneous
emission~\cite{agarwal}, and error control in quantum search algorithms
via atomic arrays~\cite{scully}.

\subsection{The general case}

The above example can be generalized to BB decoupling of a generic
(finite-dimensional) open quantum system $S$ coupled to its
environment $E$ as in Eq. (\ref{open})~\cite{vkl,vlk}. Under the
assumption that, as stipulated so far, $H_0$ is time-independent and
the applied control action is cyclic with a period $T_c$, the
stroboscopic controlled evolution of the combined system may be
described by a propagator
\begin{equation}
U (t=N T_c) =\exp(-i H_{eff} t) \:, \hspace{3mm}N \in {\mathbb N}\:,
\end{equation}
for a {\em time-independent} effective Hamiltonian $H_{eff}$. 
If, in addition, $T_c$ is sufficiently short, $H_{eff}$ is accurately 
represented by the following lowest-order {\em average Hamiltonian}:
\begin{equation}
H_{eff} \rightarrow \overline{H}^{(0)}=\frac{1}{T_c} 
\int_0^{T_c} dx \,U^\dagger_c(x) H_0 U_c(x) + \text{O}(T_c) \:.
\label{average}
\end{equation}
The term O$(T_c)$ accounts for higher-order corrections which can
be systematically evaluated through the appropriate Magnus 
expansion~\cite{ernst}, and converge to zero as the fast control
limit $T_c \rightarrow 0$ is approached. 

The basic idea of a BB protocol is to appropriately map the time
average contained in Eq. (\ref{average}) into a {\em group-theoretic}
average, by assigning the values of the control propagator $U_c(t)$
according to a discrete {\em decoupling group} ${\mathcal G}$, 
faithfully and unitarily represented on the state space ${\mathcal
H}_S$ of $S$.  ${\mathcal G}$ is determined by the set of attainable 
BB control operations.  If ${\mathcal G}=\{ g_0, g_1,\ldots, g_L\}$,
with $L=|{\mathcal G}|-1$, then $T_c$ is divided in $L-1$ equal 
subintervals of length $\Delta t$, and BB decoupling according to 
${\mathcal G}$ is implemented by sequentially sampling the 
control propagator $U_c(t)$ from ${\mathcal G}$,
\begin{equation}
U_c\Big(t=(\ell-1)\Delta t + s\Big)= g_{\ell-1}\:,
\hspace{3mm} \ell=1,\ldots,L\,,\; s\in [0,\Delta t)\:.
\label{bb}
\end{equation}
Thus, the control propagator is {\em instantaneously} changed 
from $g_{\ell-1}$ to $ g_{\ell}$ through the application of a 
BB pulse $p_\ell=g_\ell g_{\ell -1}^\dagger$ at the end of each 
control subinterval, $t_\ell=\ell \Delta t$.  In this setting, CP 
decoupling corresponds to decoupling according to the simplest
nontrivial group ${\mathbb Z}_2$, realized as ${\mathcal 
G}_{CP}=\{ {\mathbb I}, \sigma_x\}$ on ${\mathcal H}_S =
{\mathbb C}^2$, with BB control pulses $p_1=p_2=\sigma_x=
\hat{\pi}$. 

The advantage of the group-theoretic framework is that it allows
to straightforwardly construct controlled evolutions with 
well-defined symmetry properties with respect to ${\mathcal G}$,
which can in turn be exploited to filter out the unwanted 
decohering terms. In particular, it was shown in~\cite{vkl,z} 
that the effective Hamiltonian of Eq. (\ref{average}) takes 
the form
\begin{equation}
\overline{H}^{(0)}= \frac{1}{|{\mathcal G}|} 
\sum_{g_\ell \in {\mathcal G}} \, g_\ell^\dagger H_0 g_\ell \:.
\label{hsymm}
\end{equation}
Because the above Hamiltonian is invariant under ${\mathcal G}$, 
$[g_\ell , \overline{H}^{(0)}]=0$ for all $\ell$, this implies a 
controlled dynamics which is effectively {\em symmetrized}
according to ${\mathcal G}$ over any timescale longer than the 
averaging period $T_c$.  The exploitation of the symmetry 
structure enforced in this way by the controller is central to    
all aspects of the BB decoupling program, from the design of 
averaging schemes for relevant classes of environmental 
couplings~\cite{vkl}, to the characterization of universal
procedures for decoupled control~\cite{vlk}, and the 
identification of dynamically generated noise-protected 
subsystems~\cite{vkl2}. 

While being conceptually attractive thanks to the simplicity of 
the control design, BB decoupling schemes suffer from severe 
shortcomings in terms of their practical viability.  In particular,
the required control resources appear very unrealistic for two 
primary reasons: 
\begin{itemize}
\item Timescale requirements.  Reaching control timescales $T_c$
as short as the minimum correlation time of the dynamics to be 
removed may imply prohibitively {\em fast} timescales for 
realistic open quantum systems.

\item Amplitude requirements. Even in the case where the control 
period $T_c$ is allowed to be reasonably long, instantaneous 
decoupling pulses imply {\em unbounded} control strengths which
are never affordable in practice.  
\end{itemize}

Thus, the relevant question is: Can we decouple under more realistic 
assumptions?

\section{Relaxing amplitude requirements: Eulerian control design}

It turns out that the amplitude constraint can always be relaxed, at 
the expenses of making the control design slightly more sophisticated.  
The basic idea, introduced in~\cite{vk}, is to replace the piecewise 
constant BB propagator given in Eq. (\ref{bb}) with a continuous
control propagator able to ensure the same averaging, possibly 
with a reasonable overhead in the required cycle time.  This is 
accomplished by constructing the new control propagator $U_c(t)$ 
according to so-called {\em Eulerian cycles} on the Cayley graph 
$G$ of ${\mathcal G}$ that is, closed paths on $G$ which possess
the special property of using each edge of the graph once and 
once only~\cite{bollobas}.

The working principles of Eulerian control are best appreciated by
comparing Eulerian and BB implementations in the simplest case of 
decoherence suppression in a single qubit already considered in
Section 3.1.  In this case, ${\mathcal G}= {\mathcal G}_{CP}=
\{ {\mathbb I}, \sigma_x\}$, and the general BB prescription of 
Eq. (\ref{bb}) takes the explicit form
\begin{eqnarray}
\overline{\sigma_z}^{\,(BB)} &=& \frac{1}{2 \Delta t}  \Big(
\int_0^{\Delta t} ds \,{\mathbb I} \sigma_z  {\mathbb I} + 
\int_{0}^{\Delta t} ds \, {\sigma_x} \sigma_z  {\sigma_x} \Big)\nonumber \\
 &= & \frac{1}{2} \left( \sigma_z +  \sigma_x \sigma_z  \sigma_x \right)
=  \frac{1}{2} \left( \sigma_z - \sigma_z \right) =0 \:.
\label{bbb}
\end{eqnarray} 
Physically, the first $\pi$-pulse may be thought of as instantaneously 
reversing the sign of the underlying $\sigma_z$ Hamiltonian, causing the 
evolution during the second control subinterval to effectively retrace 
itself and lead to a zero average over $T_c$.  For Eulerian implementation,
control actions are no longer instantaneous, but rather distributed along 
the whole duration of each control subinterval $\Delta t$.  The relevant 
Cayley graph for ${\mathcal G}\simeq {\mathbb Z}_2$ is depicted in figure 
2.  Because there is only one generator, $\gamma=\sigma_x$, the Eulerian 
prescription turns out to involve no time overhead with respect to the BB 
case, thus $T_c=2\Delta t$ as before. Let $h_x(t)=f(t)\sigma_x$ be any 
Hamiltonian realizing the generator $\gamma$ over $\Delta t$ that is, 
\begin{equation}
u_x(t)=\text{T}\exp\left\{ -i \int_0^t ds \, h_x(s) \right\}\,, 
\; \text{ such that }\; u_x(\Delta t) = \gamma=\sigma_x \:.
\end{equation}
\begin{figure}[t]
\begin{center}
\includegraphics[width=3in,height=1.7in]{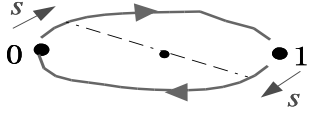}
\end{center}
\vspace*{-2cm}
\caption{ {\small Cayley graph of the group ${\mathbb Z}_2 =\{0,1\}$,
represented as ${\mathcal G} =\{ {\mathbb I}, \sigma_x \}$ on 
${\mathbb C}^2$.  Points along the path which are specularly located
relative to the center of the graph are correlated from the point of
view of the averaging process. The parameter $s\in [0,\Delta t)$ 
measures the elapsed time within each subinterval. } }
\end{figure}
On the graph pictured in figure 2, an Eulerian cycle ${\mathcal P_E}$ 
is described by the sequence of edges ${\mathcal P_E}=(\gamma, \gamma)$.
Because the path determines the sequence of control Hamiltonians to
be turned on along the cycle, in this case one simply obtains a 
control propagator given by
\begin{equation}
U_c(t) = \left\{\begin{array}{ll} 
          u_x (t) & t \in [0,\Delta t) \\
          u_x(s)\sigma_x &  t \in [\Delta t, \Delta t+s)\:, \:
s\in [0,\Delta t) \:. \end{array} \right.
\end{equation}
By inserting the above expression in Eq. (\ref{average}), the net phase 
evolution over a cycle becomes 
\begin{eqnarray}
\overline{\sigma_z}^{\,(E)} &\hspace{-1mm}=\hspace{-1mm}& 
\frac{1}{2 \Delta t}  \Big(
\int_0^{\Delta t} ds \,u_x^\dagger(s) \sigma_z  u_x (s) +
\int_{0}^{\Delta t} ds \, {\sigma_x} u_x^\dagger(s) \sigma_z u_x (s) 
{\sigma_x} \Big)\nonumber \\
 &\hspace{-1mm}=\hspace{-1mm}&   \frac{1}{2 \Delta t } \Big( 
\int_{0}^{\Delta t} ds \,u_x^\dagger(s) \sigma_z u_x(s) +  
\int_{0}^{\Delta t} ds 
\,u_x^\dagger(s) (\sigma_x \sigma_z  \sigma_x) u_x (s) 
\Big) \nonumber \\
&\hspace{-1mm}=\hspace{-1mm}&    \frac{1}{2 \Delta t }
\int_{0}^{\Delta t} ds\Big( u_x^\dagger(s) \sigma_zu_x(s) -  
 u_x^\dagger(s) \sigma_z  u_x (s) \Big) =0\:.
\label{eul}
\end{eqnarray} 
By comparing Eq. (\ref{eul}) with Eq. (\ref{bbb}), one sees that the Eulerian 
protocol still achieves averaging by inducing an effective time reversal in 
the second control subinterval; however, such a reversal is now {\em local} 
in time, the evolution in the second interval {\em continuously undoing} the 
one in the first owing to the cancellations from opposite points in the
graph (see figure 2).

The generalization of the above procedure to generic decoupling protocols
is described in~\cite{vk}.  Under mild assumptions on the relevent control 
Hamiltonians, one can always ensure that the same ${\mathcal G}$-symmetri- 
zation of the BB limit is retained.  However, Eulerian design turns out to 
be superior in two important respects: not only can any desired decoupling 
protocol be fully implemented using {\em bounded-strength} Hamiltonians, but
the resulting schemes are also intrinsically stable against a large class of
{\em systematic} control errors, thereby leading to enhanced 
robustness.

\section{Relaxing timescale requirements: 
Decoupling of low-frequency noise} 

Going back to reconsidering decoupling timescale requirements, it turns
out that they can be substantially weakened for a wide class of environments
generating low-frequency noise.  I will focus here on the most notable 
representative instance, provided by noise processes exhibiting a 1/$f$ 
spectrum.  Noise phenomena consistent with a 1/$f$ behavior have been 
identified to play a role in an astonishing variety of dynamical systems, 
ranging from physics to biology, economics, and more.  At the quantum level, 
1/$f$ noise due to a variety of material-specific fluctuation mechanisms is 
ubiquitous in solid-state devices, with a severe impact on the performance 
of both nanoelectronic circuits~\cite{fn} and qubit realizations.  In
particular, 1/$f$ noise due to fluctuating background charges is currently
regarded as one of the primary decoherence sources for superconducting 
charge qubits~\cite{naka}. 

An appropriate schematization of 1/$f$ noise effects requires going beyond
the notion of a continuum of weakly coupled environmental modes at 
equilibrium which is embodied by standard harmonic-oscillator reservoirs. 
A quantum-mechanical model capturing the distinctive features of a realistic 
{\em discrete} environments generating 1/$f$ noise was recently investigated
in~\cite{paladino}.  Within a semiclassical approach that will suffice to the 
present discussion, a simple picture of 1/$f$ noise may be obtained by 
considering an incoherent superposition of relaxation processes from an 
ensemble of $M$ independent {\em bistable fluctuators}.  Let $\xi_k(t)$ 
denote a random telegraph noise (RTN) signal describing the switching of 
an individual fluctuator between two values $\pm v_k/2$ with a total 
switching rate $\gamma_k$, and a corresponding Lorentzian power spectrum
$S_k(\omega) = \gamma_k /(\gamma_k^2 +\omega^2)$, $\omega \in {\mathbb R}$.
Provided that the switching rates in the ensemble are assumed to be 
distributed in a range $[\gamma_{min}, \gamma_{max}]$ with probability 
\begin{equation}
{\mathbb P}(\gamma) =  \frac{\text{const}}{\gamma}\,, 
\hspace{3mm}\text{const}=[\ln(\gamma_{max}/\gamma_{min})]^{-1} \:,
\end{equation}
the total fluctuation $\Xi(t) = \sum_k \xi_k(t)$ exhibits a power
spectrum $S(\omega) = \sum_k S_k (\omega)= A/|\omega|$, $A>0$, in a 
frequency interval $\gamma_{min}\ll \omega \ll \gamma_{max}$. 
For distributions of coupling strengths $\{ v_k\}$ sufficiently 
peaked around their mean value $\langle v \rangle$, the parameter 
$A$ is directly related to the product $N_d \langle v\rangle^2$, 
$n_d= M/\log(\gamma_{max}/\gamma_{min})$ being the number of 
fluctuators per noise decade. 

The simplest scenario for decoupling of 1/$f$ noise is obtained by
considering a single qubit in the presence of the semiclassical RTN 
disturbance.  Thus, the target system is described by a {\em 
time-dependent} effective Hamiltonian of the form 
\begin{equation}
H_0(t) = H_Q + H_{RTN}(t) = [\Omega \sigma_z + \Delta \sigma_x] + 
 \Xi(t) \sigma_z\:,
\end{equation}
for real parameters $\Omega, \Delta$.  While an appropriate extension
of the basic average Hamiltonian formalism is needed in general to
handle time-dependent target dynamics, the required control action can
be implemented in this case through the same CP-protocol discussed for
the spin-boson example of Section 3.1, either in the BB limit or in
the Eulerian form illustrated in the previous section.  A more
detailed analysis of this problem is available in~\cite{fv} (see
also~\cite{lidarf,gutmann,falci,galperin} for related work).  In what
follows I only focus on some illustrative results in the pure
dephasing limit, $\Delta =0$. In this regime, the stroboscopic
evolution of the qubit coherence element may be exactly evaluated,
\begin{equation}
|\rho_{01}(t=NT_c)| = \exp(-\Gamma_c (t)) \, |\rho_{01}(t=0)| \:, 
\end{equation}
the analytic expression of the controlled decoherence functional
being given in~\cite{fv}.  A formal result similar to Eq. (\ref{suppr}) 
can be explicitly established in the continuous limit where 
$T_c\rightarrow 0, N\rightarrow \infty$. Thus, {\em complete}
compensation of 1/$f$ effects requires access to control 
timescales $\Delta t \leq 1/\gamma_{max}$, so as to average out 
the influence of the {\em fastest} fluctuator present in the 
ensemble -- as expected on general control-theoretic grounds.
However, substantially slower decoupling rates might turn out to 
suffice in this case provided that the requirement of perfect 
suppression is lifted.  

\begin{figure}[tb]
\psfrag{x}{$\hspace*{-8mm} $ {\small time $(1/\Omega)$}}
\psfrag{y}{$\hspace*{-6mm} {\text{e}^{-\Gamma_c(t)}}$}
\begin{center}
\includegraphics[width=2.9in,height=3.3in]{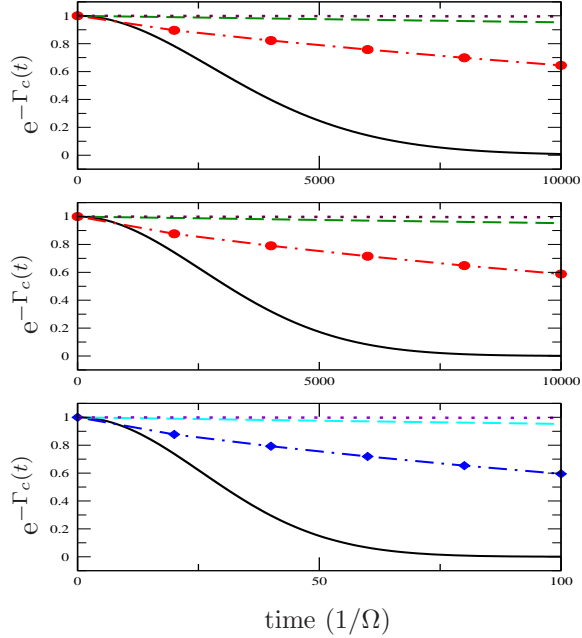}
\end{center}
\caption{ {\small Suppression of 1/$f$ pure dephasing,
$\Omega=1, \Delta=0$, starting from the initial qubit state 
$|\psi\rangle=(|0\rangle+|1\rangle)/\sqrt{2}$. 
Solid lines: free evolution.  For controlled evolutions, 
stroboscopic data (circles and diamonds) are connected by 
eye-guiding lines.  In all cases, $\gamma_{max}=100$. 
From top to bottom, the other spectral
parameters are as follows: 
(a) $\gamma_{min}=10^{-4}$, $\langle v\rangle = 10^{-4}$; 
(b) $\gamma_{min}=10^{-6}$, $\langle v\rangle = 10^{-4}$;
(c) $\gamma_{min}=10^{-4}$, $\langle v \rangle = 10^{-2}$.  
Control parameters are: 
(a) and (b) $\Delta t = 1000$ (dot dashed), 
$\Delta t = 100$ (dashed), $\Delta t = 10$ (dotted);  
(c) $\Delta t = 10$ (dot dashed), $\Delta t = 1$ (dashed), 
$\Delta t = 0.1$ (dotted).  
} }
\end{figure}

The situation is pictorially summarized in figure 3.  The
three spectra differ due to the more or less pronounced Gaussian 
character of the underlying fluctuations, purely Gaussian dephasing 
(top plot) being closest to a formal representation in terms of a 
standard oscillator bath.  As evidenced from the data, a $60\%$ coherence 
recovery is always achieved by $\Delta t$ values which are {\em at least 
three orders of magnitude longer} than expected from the inverse 
upper cutoff, the latter figure improving to five orders of 
magnitude for slower dephasing processes as in the top and 
middle panels.   While such a favorable scaling in terms of 
decoupling resources might seem in a sense to naturally follow from
the predominance of low-frequency modes in the noise profile, it is
worth stressing that similar conclusions are {\em not} immediately 
applicable to the full class of possible 1/$f$ dynamics, the actual 
decoupling performance being in practice fairly sensitive to both the 
presence of strongly non-Gaussian noise sources and the operating 
point of the qubit~\cite{fv}.  Yet, this example convincingly
shows the existence of realistic decoherence scenarios where 
decoupling might be achievable under affordable control resources. 

\section{Conclusions}

Dynamical decoupling techniques offer a well-defined conceptual framework 
for addressing a variety of open-loop coherent-control problems for quantum 
systems.  While their use as a tool for decoherence control is particularly
relevant, the applicability of decoupling methods and concepts extends to
a broader setting, encompassing schemes for quantum-dynamical engineering
on both physical and encoded degrees of freedom~\cite{v}. 

From a practical standpoint, the implementation of decoupling schemes
poses stringent requirements on the needed control operations. Although 
these are still beyond the reach of current experimental capabilities 
for many systems, progress in ultrafast coherent control is 
steady~\cite{bucks}, hopefully making 
decoupling implementations less demanding for such  systems in a near 
future.  On the theoretical front, both the notion of Eulerian control 
and the decoupling of 1/$f$ noise indicate how conclusions based on 
the simplest decoupling protocols or on general control-theoretic 
bounds may be overly pessimistic, and that significant improvement may 
still be possible in relevant situations.  It is my expectation that 
the convergence of better theoretical modeling and analysis, together 
with practical advances in quantum technologies, will improve the 
prospects that dynamical decoupling eventually becomes a method of 
choice for decoherence control.

\section*{Acknowledgments}
The research described here would not have been possible without the
contribution of many colleagues who have shared my enthusiasm in
developing and applying decoupling methods. In particular, it is a
pleasure to thank Manny Knill and Seth Lloyd for a longstanding and
fruitful collaboration, and Lara Faoro for the recent work on 1/$f$
noise suppression.  Support from the Los Alamos Office of the Director
through a J. R. Oppenheimer fellowship is gratefully acknowledged.

\bibliographystyle{unsrt}

\begin{thebibliography}{99}

\bibitem{breuer} {\sc Breuer, H.-P.}, and {\sc Petruccione, F.}, 2002, 
{\em The Theory of Open Quantum Systems}  (New York: Oxford University
Press).

\bibitem{benatti} {\em Irreversible Quantum Dynamics}, 2003, 
{\sc Benatti, F.}, and {\sc Floreanini, R.}, {\sc Eds.} 
(Heidelberg: Springer-Verlag). 

\bibitem{nc}
{\sc Nielsen, M. A.}, and {\sc Chuang, I. L.}, 2000, 
{\em Quantum Computation and Quantum Information} (Cambridge:
Cambridge University Press).

\bibitem{fort}
See, for instance, {Special Focus Issue on Experimental 
Proposals for Quantum Computation}, 2000, {\em Fortschr. Phys.}, 
{\bf 48}, Number 9-11. 

\bibitem{bm}
See {\sc Brumer, P. W.}, and {\sc Shapiro, M.}, 2003, {\em 
Principles of the Quantum Control of Molecular Processes}
(New York: Wiley \& Sons) for a general account.  An experiment 
closely approaching quantum-limited sensitivity in a nanomechanical
resonator was recently reported in {\sc LaHaye, M. D.}, {\sc Buu, O.},
{\sc Camarota, B.}, and {\sc Schwab, K. C.}, 2004, {\em Science},
{\bf 304}, 74. 

\bibitem{lv}
{\sc Lloyd, S.}, and {\sc Viola, L.}, 2002, {\em Phys. Rev. A}, 
{\bf 65}, 010101, and references therein.

\bibitem{altafini}
{\sc Altafini, C.}, 2003, {\em J. Math. Phys.}, {\bf 44}, 2357.

\bibitem{pcz}
{\sc Poyatos, J. F.}, {\sc Cirac, J. I.}, and {\sc Zoller, P.}, 1996,
{\em Phys. Rev. Lett.}, {\bf 77}, 4728.

\bibitem{dfs} 
{\sc Zanardi, P.}, and {\sc Rasetti, M.}, 1997, {\em  Phys. Rev. Lett.}, 
{\bf 79}, 3306; {\sc Lidar, D. A.}, {\sc Chuang, I. L.}, and {\sc 
Whaley, K. B.}, 1998, {\em  ibid.}, {\bf 81}, 2594.

\bibitem{klv}
{\sc Knill, E.}, {\sc Laflamme, R.}, and {\sc Viola, L.}, 2000,
{\em Phys. Rev. Lett.}, {\bf 84}, 2525.

\bibitem{kita}
{\sc Kitaev, A.}, 2003, {\em Ann. Phys.}, {\bf 303}, 2; 
{\sc Bravy, S.}, and {\sc Kitaev, A.}, 1998, quant-ph/9811052.

\bibitem{zl}
{\sc Zanardi, P.}, and {\sc Lloyd, S.}, 2003,  
{\em Phys. Rev. Lett.}, {\bf 90}, 067902.

\bibitem{wise}
{\sc Wiseman, H. M.}, 1994, {\em  Phys. Rev. A}, {\bf 49}, 2133; {\em
ibid.}, {\bf 50}, 4428.  See also {\sc Ahn, C.}, {\sc  Wiseman, H. M.}, 
and {\sc Milburn, G. J.}, 2003,  {\em  Phys. Rev. A}, {\bf 67}, 052310.

\bibitem{shor}
{\sc Shor, P. W.}, 1995, {\em Phys. Rev. A}, {\bf 52}, 2493.

\bibitem{steane}
{\sc Steane, A. M.}, 1996, {\em Phys. Rev. Lett.}, {\bf 77}, 793.

\bibitem{vl}
{\sc Viola, L.}, and {\sc Lloyd, S.}, 1998,
{\em Phys. Rev. A}, {\bf 58}, 2733.

\bibitem{vkl}
{\sc Viola, L.}, {\sc Knill, E.}, and {\sc Lloyd, S.}, 1999, 
{\em Phys. Rev. Lett.}, {\bf 82}, 2417.

\bibitem{hahn} {\sc Hahn, E. L.}, 1950, {\em Phys. Rev.}, {\bf 80}, 580.

\bibitem{haeberlen} {\sc Haeberlen, U.}, and {\sc Waugh, J. S.}, 1968, 
 {\em Phys. Rev.}, {\bf 175}, 453.

\bibitem{vlk}
{\sc Viola, L.}, {\sc Lloyd, S.}, and {\sc Knill, E.}, 1999, 
{\em Phys. Rev. Lett.}, {\bf 83}, 4888.

\bibitem{z}
{\sc Zanardi, P.}, 1999, {\em Phys. Lett. A}, {\bf 258}, 77.

\bibitem{vt}
{\sc Vitali, D.}, and {\sc Tombesi, P.}, 1999, {\em Phys. Rev. A},
{\bf 59}, 4178. 

\bibitem{v}
{\sc Viola, L.}, 2002,
{\em Phys. Rev. A}, {\bf 66}, 012307, and references therein.

\bibitem{lw}
{\sc Lidar, D. A.}, and {\sc Wu, W.-A.}, 2002, 
{\em Phys. Rev. Lett.}, {\bf 88}, 017905; {\sc Wu, W.-A.}, and 
{\sc Lidar, D. A.}, 2002, {\em ibid.}, 207902. 

\bibitem{bl}
{\sc Byrd, M. S.}, and {\sc Lidar, D. A.}, 2002, 
{\em Quantum Inf. Proc.}, {\bf 1}, 19.
 
\bibitem{vk}
{\sc Viola, L.}, and {\sc Knill, E.}, 2003,
{\em Phys. Rev. Lett.}, {\bf 90}, 037901.

\bibitem{simu}
See, for instance, 
{\sc Wocjan, P.}, {\sc R\"otteler, M.}, {\sc Janzing, D.}, and 
{\sc Beth, Th.}, 2002, {\em Quantum Inf. Comput.}, {\bf 2}, 133;
{\sc Bennett, C. H.}, {\em et al.}, 2002, {\em Phys. Rev. A}, 
{\bf 66}, 012305.

\bibitem{ernst}
{\sc Ernst, R. R.}, {\sc Bodenhausen, G.}, and {\sc Wokaun, A.}, 
1994, {\em Principles of Nuclear Magnetic Resonance in One and Two
Dimensions} (Oxford: Oxford University Press).

\bibitem{speranta}
{\sc Gheorghiu-Svirchevski, S.}, 2002, {\em  Phys. Rev. A}, {\bf 66}, 
032101.

\bibitem{chi}
{\sc Uchiyama, C.}, and {\sc Aihara, M.}, 2002, {\em Phys. Rev. A}, 
{\bf 66}, 032313.

\bibitem{berglund}
{\sc Berglund, A. J.}, 2000, quant-ph/0010001.

\bibitem{search}
{\sc Search, C.}, and {\sc Berman, P. R.}, 2000, {\em  Phys. Rev. Lett.}, 
{\bf 85}, 2272.

\bibitem{agarwal}
{\sc Agarwal, G. S.}, {\sc Scully, M. O.}, and {\sc Walther, W.}, 
2001, {\em Phys. Rev. Lett.}, {\bf 86}, 4271.

\bibitem{scully}
{\sc Scully, M. O.}, and {\sc Zubairy, M. S.}, 2001, 
{\em Phys. Rev. A}, {\bf 64}, 022304.

\bibitem{vkl2}
{\sc Viola, L.}, {\sc Knill, E.}, and {\sc Lloyd, S.}, 2000, 
{\em Phys. Rev. Lett.}, {\bf 85}, 3520.

\bibitem{bollobas}
{\sc Bollob\'as, B.}, 1998, {\em Modern Graph Theory} (New York: 
Sprin- ger-Verlag).

\bibitem{fn}
{\sc Peters, M. G.}, {\sc  Dijkhuis, J. I.}, and {\sc Molenkamp, L. W.},
1999, {\em J. Appl. Phys.}, {\bf 86}, 1523; {\sc Collins, P. G.}, 
{\sc Fuhrer, M. S.}, and {\sc Zettl, A.}, 2000, {\em ibid.}, {\bf 
76}, 894; {\sc Covington, M.}, {\sc Keller, M. W.}, {\sc Kautza, R. L.},
and {\sc Martinis, J.}, 2000, {\em Phys. Rev. Lett.}, {\bf 84}, 5192.

\bibitem{naka}
{\sc Nakamura, Y.}, {\sc Pashkin, Yu. A.}, {\sc Yamamoto, T.}, and
{\sc Tsai, J. S.}, 2002, {\em Phys. Rev. Lett.}, {\bf 88}, 047901; 2002,
{\em Phys. Scripta}, {\bf 102}, 155.

\bibitem{paladino}
{\sc Paladino, E.}, {\sc Faoro, L.}, {\sc Falci, G.}, and 
{\sc Fazio, R.}, 2002, {\em Phys. Rev. Lett.}, {\bf 88}, 228304.

\bibitem{fv}
{\sc Faoro, L.}, and {\sc Viola, L.}, 2004,
{\em Phys. Rev. Lett.}, {\bf 92}, 117905.

\bibitem{lidarf}
{\sc Shiokawa, K.}, and {\sc Lidar, D. A.}, 2004, {\em Phys. Rev. A}, 
{\bf 69}, 030302.

\bibitem{gutmann}
{\sc Gutmann, H.}, {\sc Wilhelm, F. K.}, {\sc Kaminsky, W. M.}, 
and {\sc Lloyd, S.}, 2003, cond-mat/0308107. 

\bibitem{falci} 
{\sc Falci, G.}, {\sc D'Arrigo, A.}, {\sc Mastellone, A.}, and
{\sc Paladino, E.}, 2003, cond-mat/0312442.

\bibitem{galperin}
{\sc Galperin, Y. M.}, {\sc Altshuler, B. L.}, and {\sc Shantsev, 
D. V.}, 2003, cond-mat/0312490. 

\bibitem{bucks}
See, for instance, {\sc Bucksbaum, P. H.}, 2003, 
{\em Nature}, {\bf  421}, 593 for a recent highlight.

\end{thebibliography}

\end{document}